\documentclass[aps,prb,10pt,twocolumn,superscriptaddress,floatfix]{revtex4-2}
\usepackage[colorlinks, linkcolor=blue, citecolor=blue, urlcolor=blue]{hyperref}
\usepackage[dvipsnames]{xcolor}
\usepackage{graphicx}
\usepackage{dcolumn}
\usepackage{amsmath}
\usepackage{mathrsfs}
\usepackage{amssymb}
\usepackage{physics}
\usepackage{braket}

\newcommand\numberthis[1][]{%
    \refstepcounter{equation}%
    \ifx#1\empty\else\label{eq:#1}\fi%
    \tag{\theequation}%
}
\DeclareUnicodeCharacter{00A4}{\textcurrency}
\setcitestyle{numbers,square}
\begin{document}

\title{Thermoelastic Damping Across the Phase Transition in van der Waals Magnets}
\author{Alvaro Bermejillo-Seco}
\thanks{Equal contribution}
\email{a.bermejilloseco@tudelft.nl}
\affiliation{%
Kavli Institute of Nanoscience, Delft University of Technology, Lorentzweg 1,2628 CJ Delft, Netherlands
}
\author{Xiang Zhang}
\thanks{Equal contribution}
\email{xiangzhang@swu.edu.cn}
\affiliation{%
School of Physical Science and Technology, Southwest University, Chongqing 400715, China
}
\affiliation{%
Kavli Institute of Nanoscience, Delft University of Technology, Lorentzweg 1,2628 CJ Delft, Netherlands
}
\author{Maurits J.A. Houmes}
\affiliation{%
Kavli Institute of Nanoscience, Delft University of Technology, Lorentzweg 1,2628 CJ Delft, Netherlands
}
\author{Makars Šiškins}
\affiliation{%
Kavli Institute of Nanoscience, Delft University of Technology, Lorentzweg 1,2628 CJ Delft, Netherlands
}
\affiliation{%
Institute for Functional Intelligent Materials, National University of Singapore, 117544, Singapore
}
\author{Herre S.J. van der Zant}
\affiliation{%
Kavli Institute of Nanoscience, Delft University of Technology, Lorentzweg 1,2628 CJ Delft, Netherlands
}
\author{Peter G. Steeneken}
\affiliation{%
Department of Precision and Microsystems Engineering, Delft University of Technology, Mekelweg 2, 2628 CD Delft, Netherlands
}
\author{Yaroslav M. Blanter}
\email{y.m.blanter@tudelft.nl}
\affiliation{%
Kavli Institute of Nanoscience, Delft University of Technology, Lorentzweg 1,2628 CJ Delft, Netherlands
}

\date{\today}

\begin{abstract}
A quantitative understanding of the microscopic mechanisms responsible for damping in van der Waals nanomechanical resonators remains elusive. In this work, we investigate van der Waals magnets, where the thermal expansion coefficient exhibits an anomaly at the magnetic phase transition due to magnetoelastic coupling. Thermal expansion mediates the coupling between mechanical strain and heat flow and determines the strength of thermoelastic damping (TED). Consequently, variations in the thermal expansion coefficient are reflected directly in TED, motivating our focus on this mechanism. We extend existing TED models to incorporate anisotropic thermal conduction, a critical property of van der Waals materials. By combining the thermodynamic properties of the resonator material with the anisotropic TED model, we examine dissipation as a function of temperature. Our findings reveal a pronounced impact of the phase transition on dissipation, along with transitions between distinct dissipation regimes controlled by geometry and the relative contributions of in-plane and out-of-plane thermal conductivity. These regimes are characterized by the resonant interplay between strain and in-plane or through-plane heat propagation. To validate our theory, we compare it to experimental data of the temperature-dependent mechanical resonances of FePS$_3$ resonators. 

\end{abstract}

\maketitle

\section{Introduction}
\label{sec:introduction}
The discovery of van der Waals (vdW) magnets has opened new avenues for studying magnetism in the two-dimensional limit \cite{Gong2017, Huang2017}. Since this discovery, rapid advancements have identified new vdW magnetic materials, such as MPX$_3$ (with M = Fe, Co, Mn and X=S, Se) \cite{Burch2018}, they are candidates for exploring unusual phenomena in 2D magnetism \cite{Haldane1988,Balents2010}, and realized potential applications like Magnetic Tunnel Junctions (MTJs) \cite{Song2018}. An emergent area of research within vdW magnets focuses on the coupling between vibrational and magnetic degrees of freedom, known as magnetostriction \cite{Jiang2020,Houmes2023, Fei2024,Kremer2024,Bae2024}. Magnetostriction is a well-studied phenomenon in solid-state physics which affects the damping of spin waves, is crucial for spintronics \cite{Bozhko2020}, and offers potential for efficient and coherent transduction between the optical and microwave domains, a requirement for the development of quantum networks \cite{Tabuchi2015, Jie2021, Engelhardt2022}.

One promising approach to study magnetostriction in vdW magnets involves (nano)mechanical resonators. Here, changes in the magnetization in suspended vdW magnets induce variations in tension, which in turn alter the resonant frequency. In general, changes in the resonant frequency are small. However, they become more pronounced in two specific cases. First, when the material undergoes a spin reorientation phenomenon due to an external magnetic field, such as a spin flip, the resonant frequency jumps abruptly \cite{Jiang2020}. Second, when the material undergoes a magnetic phase transition, magnetostriction induces a big change in the thermal expansion due to the emergence of magnetic order and provides a clear shift in the resonant frequency of the mechanical resonator, which can be used to detect the Néel temperature~\cite{siskins2020}. Additionally, in-plane anisotropy in such nanomechanical resonators reveals information about the temperature-dependent order parameter, providing insight into the nature of magnetism in the studied material \cite{Houmes2023}. 

Besides changes in resonance frequency, the system’s dissipation, quantified by the quality factor, plays a crucial role. Theoretically investigating dissipation mechanisms in these devices is complex, since the isolation of contributions from competing sources is challenging. Considerable effort has been made to uncover these mechanisms due to the insight they provide into the resonator's physics and their importance in applications \cite{Steeneken2021, Bachtold2022}.

Numerous dissipation mechanisms affect nanomechanical resonators, such as friction with a medium or clamping losses. Here, we will focus on mechanisms related to the magnetism in the system. We identify two, thermoelastic damping and magnetoelastic damping. The latter is a direct coupling between the mechanical and spin wave modes. We neglect this effect due to the large mismatch between the frequencies of the mechanical vibrations (typically in the MHz regime) and those of spin
waves (typically in the GHz-THz regime), which makes the coupling small. It is worth noting that although we neglect this effect in the present work it can become important for non-linear damping close to the phase transition \cite{siskins2023}. Additionally, such interaction has been studied in a different context, most notably in yttrium iron garnet (YIG), where coherent excitations enable a stronger coupling between the modes \cite{zhang2016}. 

On the other hand, thermoelastic damping considers the coupling between mechanical and thermal modes, which depending on the size of the system can be resonant as we will show later. In addition, experiments in FePS$_3$ \cite{siskins2020} and CrSBr \cite{Fei2024} have hinted that thermoelastic damping (TED) is a reasonable candidate for dissipation around the phase transition in such systems. However, a quantitative analysis of how TED is affected by such a magnetic phase transition is still lacking. Other dissipation mechanisms are listed with more detail in Table \ref{tab:dissipation} in Appendix \ref{appendix:dissipation}.

Thermoelastic damping relies on the coupling between the strain and temperature fields in the resonator mediated by thermal expansion. The time-dependent inhomogeneous strain field generates thermal gradients across the resonator, leading to irreversible heat flow. Modeling TED has been essential for the development of ultra-high quality factor resonators, with significant progress made since the first description by Zener \cite{zener1937}, and later detailed for one-dimensional systems by Lifshitz and Roukes \cite{Lifshitz2000}. Both works focused exclusively on out-of-plane thermal conduction, as the thermal gradient mostly relaxes along that direction due to the high aspect ratio of the resonators. Prabhakar and Vanglatore \cite{Prabhakar2008} quantified the error of this approximation and extended the model to include two-dimensional heat conduction. This extension has been further explored \cite{Ma2020,Yongpeng2020,zhou2023}, but to our knowledge only for materials with isotropic thermal conductivity $\kappa$. In contrast, vdW materials are known to be highly anisotropic, with a ratio of in-plane ($\parallel$) to out-of-plane ($\perp$) thermal conductivity, $\kappa$, reaching up to 900 in MoS$_2$ in the presence of stacking defects \cite{Kim2021}. For FePS$_3$, this ratio $\kappa_\parallel/\kappa_\perp$ has been reported to be 3.2 at room temperature, with $\kappa_\parallel=2.7$ W/Km and $\kappa_\perp=0.85$ W/Km, and for MnPS$_3$ the ratio is 5.7 \cite{Kargar2020}. Based on this important property of van der Waals materials, we develop an anisotropic two-dimensional thermal conduction model for TED and investigate to what extent the in-plane versus out-of-plane anisotropy affects dissipation compared to typical nanomechanical resonators made with isotropic materials like Si or SiN.

In this Article, we propose a model to predict the effect of thermoelastic damping in suspended anisotropic vdW magnets. First, we describe the system in detail in Section \ref{sec:system}, where we choose the material and geometry to use as an example. Then, in Section \ref{sec:model} we extend the existing TED models to incorporate anisotropic thermal conduction, which is ubiquitous in vdW materials. The results of Section \ref{sec:model} apply to all anisotropic 2D materials. For the rest of the Article, we concentrate on magnetic materials that undergo a phase transition. In Section \ref{sec:thermal} we study the thermodynamical properties of the material and show how they are affected by magnetostriction. In Section \ref{sec:results}, we use the model to predict the temperature and radius dependence and compare it with experimental measurements. Finally, in Section \ref{sec:discussion} we discuss the results and come to some conclusions. 

\section{System description}
\label{sec:system}
\begin{figure}[ht]
\includegraphics[width=0.45\textwidth]{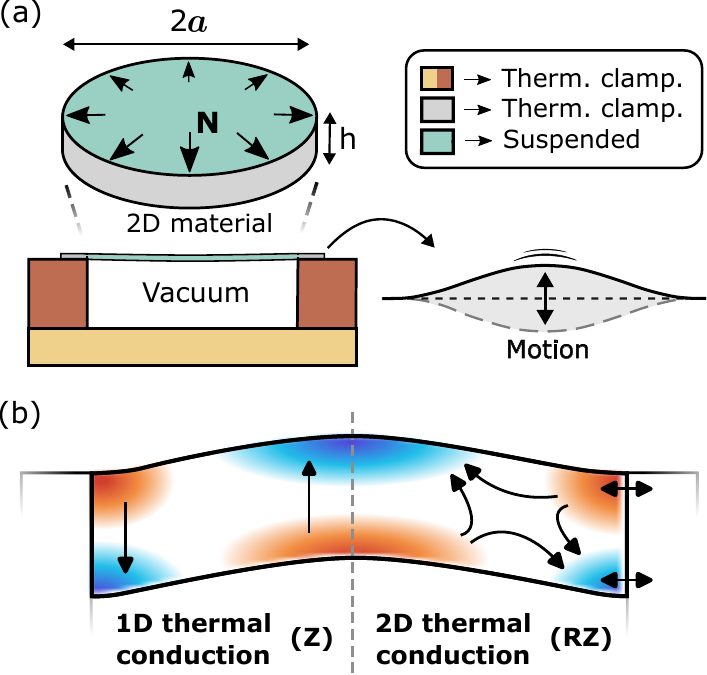}
\caption{Description of the system. \textbf{a)} A van der Waals magnet is suspended over a cavity. The 2D magnetic material forms a thin plate of thickness $h$ and radius $a$, which can oscillate freely in the out-of-plane direction. The system is in vacuum, and the temperature is controlled by thermalizing the substrate. As the temperature is changed, a uniform radial tension, N, is exerted on the suspended material. \textbf{b)} Schematic representation of thermoelastic damping. The bending generates compressed regions (heated, in red) and expanded regions (cooled, in blue), such that heat currents propagate through the material giving rise to mechanical energy losses. Two models for heat transport are represented: an approximation considering only through-plane heat conduction (Z-model) and one including in-plane heat conduction in the radial direction (RZ-model).}
\label{fig:fig1}
\end{figure}

The system under study is described in Figure \ref{fig:fig1}. It consists of a thin magnet suspended over a cavity etched in a substrate (e.g., Si/SiO\textsubscript{2}), as depicted in panel a), to which it is clamped via van der Waals forces. We choose a plate with circular geometry of radius $a$ and thickness $h \ll a$. The suspended material can then oscillate freely in the out-of-plane direction. As the temperature changes, a tension, $N$, develops in the suspended material due to the mismatch in the thermal expansion coefficients of the suspended material and the substrate. 

Due to thermoelasticity, when different areas of the suspended material are compressed (expanded), they heat up (cool down) locally. The temperature gradients between these regions lead to heat currents which cause TED. This is depicted in Figure \ref{fig:fig1} b) with blue areas showing cold (expanded) regions and red areas showing hot (compressed) regions. Typically an assumption is made that only the temperature gradients across the small thickness contribute, due to the high aspect ratio between thickness (tens of nanometers) and lateral size (several micrometers) in typical nanomechanical resonators \cite{zener1937, Lifshitz2000}. We call the model considering only the through-plane thermal conduction the Z-model and the model where we drop that assumption the RZ-model, in which in-plane heat conduction is also considered. In panel b) we represent these models with different thermal conduction conditions that are discussed throughout this paper. We highlight that the only anisotropy considered in this work is related to the thermal conductivity, and for instance, the system is regarded as isotropic in terms of its elastic properties. 

\section{Thermoelastic damping with anisotropic thermal conductivity}
\label{sec:model}

Thermoelastic damping is described by the dissipation factor $\mathcal{Q}^{-1}$ which is defined to be the ratio of energy loss per cycle to the energy stored in the oscillator~\cite{Lifshitz2000},
\begin{equation}
    \mathcal{Q}^{-1}=\frac{1}{2\pi}\frac{\text{energy lost per cycle}}{\text{energy stored}},
\end{equation}
where $\mathcal{Q}$ is the mechanical quality factor of the resonator. To obtain the dissipation in the system, first, we model the thermoelastic interaction in the system in Subsection \ref{subsec:model} and we use the obtained results to compute the energy lost per cycle and stored energy in Subsection \ref{subsec:dissipation}. 

\subsection{Model of thermoelasticity}
\label{subsec:model}

To model the system we assume that the suspended material behaves as a clamped thin plate with isotropic elastic properties. The equation of motion for the displacement of the middle surface, $w$, of a thin plate with uniform radial tension $N$, is given by~\cite{leissa2011}
\begin{equation}
    D\nabla^4w - N\nabla^2 w + \rho h \frac{\partial^2 w}{\partial t^2}=0.
    \label{eq:plate}
\end{equation}
Here, $D = Eh^3 / \big(12(1-\nu^2)\big)$ is the bending rigidity -- which depends on the Young's modulus, $E$, and the Poisson ratio, $\nu$, -- and $\rho$ is the mass density of the material. We consider harmonic oscillations of the plate such that $w=W(r,\theta)e^{i\omega t}$ with $\omega$ the frequency of the mechanical oscillation and $W$ describes the mode shape. It is described in cylindrical coordinates, with $r$, $\theta$ and $t$, the radial, angular, and temporal coordinates.  The solution for the fundamental mode of Eq. (\ref{eq:plate}), and its associated resonant frequency, are given by \cite{Ma2020}
\begin{equation}
\begin{split}
     W(r) &=  J_0\left(\frac{\delta_- r}{a}\right)-\frac{J_0(\delta_-)}{I_0(\delta_+)}I_0\left(\frac{\delta_+ r}{a}\right), \\
     \omega &= \frac{\sqrt{\left(\delta_-^2 +\delta_+^2\right)^2-\left(a^2 N /D\right)^2}}{2a^2}\sqrt{\frac{D}{\rho h}}.
\end{split}
    \label{eq:plate_sol}
\end{equation}
Here, $J_0$ and $I_0$ are the zeroth order Bessel and modified Bessel function of first kind, and $\delta_-$ and $\delta_+$ are two parameters. These parameters modulate the weight between $J_0$ and $I_0$ for the plate's shape depending on the tension N and are found by simultaneously solving two equations shown in Eq. (\ref{eq:deltas}) of Appendix~\ref{appendix:temperature}adial tension is given by the mismatch of the thermal expansion of the membrane and the substrate as the temperature of the system is changed. Thus, the tension in the plate is given by
\begin{equation}
    N(T) = \frac{Eh}{1-\nu}\int_{T_0}^{T}(\alpha_{\text{membrane}}-\alpha_{\text{substrate}})dT,
    \label{eq:tension}
\end{equation}
where $T_0$ stands for room temperature, and $\alpha$ is the isothermal thermal expansion coefficient.  In a real system, the total tension has additional contributions, such as pre-strain, $N_0$, induced in fabrication, and the thermally generated tension: $N_{\text{tot}}(T)=N_0+N(T)$. 
\begin{figure}[htb]
\includegraphics[width=0.45\textwidth]{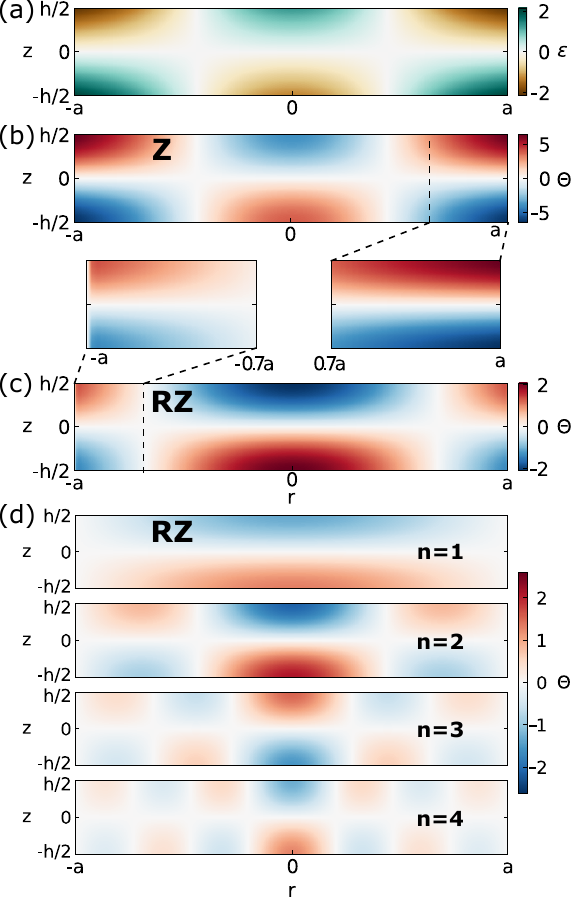}
\caption{Strain and temperature profiles (arbitrary units).
     \textbf{a)} Volumetric strain profile of a plate.
     \textbf{b)} Temperature profile according to the Z-model, with the volumetric strain profile in a) as a heat source.
     \textbf{c)} Temperature profile according to the RZ-model, with the volumetric strain profile in a) as a heat source. The extra boundary condition significantly changes the heat mode shape at the edges, as seen in the zoomed-in cuts. Additionally, the central regions are more extended with respect to the Z-model, and the temperature deviation is reduced to, approximately, half.
     \textbf{d)} Individual modes of the temperature profile in the RZ-model. In the RZ-model, the temperature profile is an infinite sum of modes dependent on the zeros of $J_0$, $j_0^m$. The second mode has the most amplitude due to the big overlap with the strain profile shown in \textbf{a)}.}
\label{fig:fig2}
\end{figure}

To compute the temperature field generated by the motion described by $w(r,t)$, we first need to find the strain profile in the plate. The strain components can be divided into two different contributions, namely, the strain induced by tension ($\epsilon^N$) and the oscillating flexural strain caused by the vibrations ($\epsilon^{\text{flex}}$):
\begin{equation}
    \epsilon_i(r,z,t) = \epsilon^{\text{flex}}_i(r,z)\cdot e^{i\omega t}+\epsilon_i^N, \qquad i=(r,\theta, z).
    \label{eq:strain}
\end{equation}
These components are directly related to the mid-plane deflection, $W$, such that the different contributions take the following forms \cite{leissa2011}
\begin{equation}
\begin{split}
    &\epsilon^{\text{flex}}_r(r,z)=-z\frac{\partial^2 W}{\partial r^2}, \quad \epsilon^{\text{flex}}_\theta(r,z)=-\frac{z}{r}\frac{\partial W}{\partial r}, \\ &\epsilon_r^N(r,z)=\epsilon_\theta^N(r,z)=\frac{(1-\nu)N}{Eh}.
\end{split}
    \label{eq:straincomp}
\end{equation}
The shear term in the $r$,$\theta$ direction is zero for axisymmetric modes, and the $z$ components of the strain are zero throughout the whole plate due to Kirchhoff's assumption, valid for small deflections of thin plates \cite{LandauLifshitz}. 

We obtain the generated temperature distribution in the plate via the one-way coupled thermoelastic equation \cite{Prabhakar2008}
\begin{equation}
    \rho c_\text{V} \frac{\partial T}{\partial t} + \beta_{\text{TE}} T \frac{\partial \epsilon}{\partial t}=\kappa \nabla^2T,
    \label{eq:thermoelastic}
\end{equation}
where $c_\text{V}$ is the specific heat, $\beta_{TE}$ the thermoelastic coupling given by $\beta_{\text{TE}}=\alpha_\text{T} E/(1-2\nu)$ and $\epsilon=\delta V/V$ stands for the volumetric strain; here $\epsilon=\epsilon_r+\epsilon_{\phi}+\epsilon_z$ in cylindrical coordinates. Considering that the oscillations of the system are small and that thermoelastic coupling is weak \cite{Day1985}, we can linearize Eq. (\ref{eq:thermoelastic}). To do so we consider a small temperature field, $\Theta$, such that the temperature in the system is given by
$T(r,t)=T_0+\Theta(r,t)$ where $T_0$ is the environment temperature and $\Theta$ stands for the temperature field within the material. Then, Eq. (\ref{eq:thermoelastic}) reads
\begin{equation}
    \rho c_\text{V} \frac{\partial \Theta}{\partial t} + \beta_{\text{TE}}(T_0+\Theta) \frac{\partial \epsilon}{\partial t}=\kappa \nabla^2\Theta,
\end{equation}
in which we can neglect the second-order term $\beta_{\text{TE}}\Theta \frac{\partial \epsilon}{\partial t}$.
Including explicitly different in-plane and out-of-plane thermal conductivities, it reads
\begin{equation}
\begin{split}
    \rho c_\text{V} \frac{\partial \Theta}{\partial t}= &\left[\kappa_\perp\frac{\partial^2}{\partial z^2} + \kappa_\parallel\left(\frac{\partial^2}{\partial r^2} + \frac{1}{r}\frac{\partial}{\partial r}\right)\right] \Theta \\
    &-\beta_{\text{TE}} T_0 \frac{\partial \epsilon}{\partial t}.
\end{split}
\label{eq:anis_thermoelastic}
\end{equation}
Using $\Theta=\Theta_0e^{i\omega t}$, considering no thermal exchange at the boundaries $\partial \Theta_0 /\left.\partial z\right|_{z= \pm h / 2}=0$ and assuming perfect thermal contact at the edges $\left.\Theta_0(r, z)\right|_{r=a}=0$, the solution to the temperature field is given by
\begin{equation}
    \Theta_0(r,z)=\sum_{n=1}^\infty J_0\left(\frac{j_0^n}{a}r\right)\frac{c_n}{b_n^2}\left[\frac{1}{b_n}\frac{\sinh(b_nz)}{\cosh(b_nh/2)}-z\right].
    \label{eq:sol2d}
\end{equation}
Here, $j_0^n$ are the zeros of $J_0$, and $c_n$ and $b_n$ are complex-valued parameters shown in Appendix~\ref{appendix:solving}.

Fig. \ref{fig:fig2} shows the strain profile and temperature distributions calculated considering only $\kappa_\perp$ (Z-model) and both $\kappa_\perp$ and $\kappa_\parallel$ (RZ-model). The calculation corresponds to a 45 nm thick, 5 $\mu$m radius plate with an amplitude of 1 nm, and material parameters corresponding to FePS$_3$ at 200K. The temperature profile in the Z-model (panel b) follows exactly the strain profile (panel a), with the strongest temperature gradients at the edges of the membrane. In contrast, in the RZ-model, due to the addition of the perfect thermal contact boundary condition at $r=a$, the temperature gradients there are smaller than in the center of the membrane. The shape of the temperature gradient does not follow the strain profile anymore, as it expands in the in-plane direction due to the in-plane thermal conduction. In panel d) we can see the first 4 thermal modes contributing to the temperature profile. 

Beyond modifying the resulting strength of thermoelastic damping, this mismatch between the strain and the temperature profiles could lead to intermode coupling. As higher modes of the temperature profile become occupied, they interact with higher-order mechanical modes due to the non-zero overlap of temperature and strain. Contrarily, this is not allowed in the Z model, where the lack of conductivity in the plane prevents other thermal modes from being excited. In this article, we concentrate on the first mode, but it would be interesting to explicitly address this effect in future research.

\subsection{Calculation of the dissipation}
\label{subsec:dissipation}

Once this complex-valued temperature field is known, the dissipated energy per cycle can be computed by evaluating the work done by the external stress on the total generated strain \cite{Schmid2016}, which is given by
\begin{equation}
   \Delta E = \oint\epsilon d\sigma = \int_VdV\int_0^{2\pi/\omega}\epsilon\frac{d\sigma}{dt}dt.
   \label{eq:deltaE}
\end{equation}

The real part of the mean displacement-induced stress, $\sigma$, for an isotropic plate reads
\begin{equation}
    \sigma = \frac{E}{1-\nu}\epsilon = \frac{E}{1-\nu}(\epsilon^{\text{flex}}\cos(\omega t) + \epsilon^N),
\end{equation}
and its time derivative
\begin{equation}
    \frac{d\sigma}{dt} = \omega \frac{E}{1-\nu}\epsilon^{\text{flex}}\sin(\omega t).
    \label{eq:dsigmadt}
\end{equation}
The real part of the total strain generated by the stress also has the thermal contribution generated in the thermoelastic process coming from the temperature field computed before
\begin{equation}
    \epsilon_{\text{tot}} = (\epsilon^{\text{flex}}+\Re\{\epsilon^{\text{th}}\})\cos(\omega t)+\Im\{\epsilon^{\text{th}}\}\sin(\omega t)+\epsilon^N,
    \label{eq:epsilon_tot}
\end{equation}
where the volumetric thermal strain is $\epsilon^{\text{th}} =3\alpha_\text{T}\Theta_0$. The prefactor 3 comes from the relation between volumetric and longitudinal thermal expansion coefficients. Introducing Eq. (\ref{eq:dsigmadt}) and (\ref{eq:epsilon_tot}) into Eq. (\ref{eq:deltaE}), we observe that due to the time integral the only non-zero component is the one proportional to $\sin^2(\omega t)$:
\begin{equation}
\begin{split}
    \Delta E &= -E\omega\int_VdV\int_0^{2\pi/\omega}\epsilon^{\text{flex}}~\Im\{\epsilon_{\text{th}}\}\sin^2(\omega t)dt\\
    &=-3\pi E\alpha_{\text{T}}\int_VdV\epsilon^{\text{flex}}~\Im\{\Theta_0\}.
\end{split}
\label{eq:DeltaEint}
\end{equation}
This is a common result for anelastic relaxation processes \cite{Nowick1972}, where the energy loss comes from the interaction between the flexural strain and the out-of-phase component of the thermal strain. The heat generated in phase with the strain gets reabsorbed by the strain field. The energy loss per cycle, $\Delta E$ is found by inserting Eq. (\ref{eq:plate_sol}), Eq. (\ref{eq:straincomp}) and Eq. (\ref{eq:sol2d}) into Eq. (\ref{eq:DeltaEint}) -- which involves finding the imaginary part and performing the integral over the volume, 
\begin{equation}
    \Delta E= -\frac{4\omega \pi^2 E^2 \alpha_\text{T}^2 T_0}{\kappa (1-2\nu)(1-\nu)a^2}\sum_{n=1}^{\infty} S_n,
    \label{eq:deltaEsum}
\end{equation}
where $S_n$ can be expressed as
\begin{equation}
    S_n=\frac{B_nh^3/12-C_n}{D_n}\left(I_n\right)^2.
    \label{eq:sum_terms_diss}
\end{equation}
The factors $B_n$, $C_n$, and $D_n$ contain information about the interplay of the mechanical frequency and the ratio of out-of-plane and in-plane thermal conductivities, and $I_n$ is a geometry factor dependent on the modeshape, W. These factors are defined in Appendix~\ref{appendix:diss_energy} Eq. (\ref{eq:factors}).

The total stored energy can be computed from the elastic potential energy at maximum deflection
\begin{equation}
     E_{\text{tot}}=\frac{1}{2}\omega^2\rho\int_VW^2dV,
     \label{eq:emax}
\end{equation}
and the mode shape of Eq. (\ref{eq:plate_sol}), which results in 
\begin{equation}
\begin{split}
    E_\text{tot}=\frac{1}{2}\pi h \rho \omega^2 a^2 \left[J_0^2(\delta_-)\left(2-\frac{4\delta_+I_1(\delta_+)}{(\delta_-^2+\delta_+^2)I_0(\delta_+)}\right.\right. \\
    \left.\left.-\frac{I_1^2(\delta_+)}{I_0^2(\delta_+)}\right)-\frac{4\delta_-J_0(\delta_-)J_1(\delta_-)}{\delta_-^2+\delta_+^2}+J_1^2(\delta_-)\right].
\end{split}
\end{equation}
The thermoelastic damping factor for this oscillation system then becomes
\begin{equation}
    Q^{-1}=\frac{1}{2\pi}\frac{\Delta E}{E_\text{tot}}\propto \Delta_\text{E}=\frac{E\alpha_T^2(1+\nu)}{\rho c_v (1-2\nu)}.
    \label{eq:dissipation}
\end{equation}
The proportionality constant $\Delta_E$ is a dimensionless quantity called the relaxation strength. It is the same constant found by \cite{zener1937} and \cite{Lifshitz2000} except for the factor related to Poisson's ratio, which appears due to the circular geometry, and is identical to that found by \cite{Ma2020}. The rest of the contribution to $Q^{-1}$ is different from all previous work due to anisotropic thermal conduction and is given by the interplay of mechanical frequency and thermal relaxation of the temperature modes, as well as geometry factors and the tension contribution.

As seen before, the factor $\mathcal{Q}^{-1}/\Delta_E$ is complicated as it entails the sum of the interactions between all the thermal modes and the strain generated in the plate for its fundamental mode. However, as we will see in the Results section, it is simple to understand qualitatively in terms of a resonance behavior between the mechanical time constant, $\tau_{\text{mech}}=2\pi/\omega$, and the thermal relaxation time constants in the $z$-direction and $r$-direction are \cite{baglioni2023}
\begin{equation}
    \tau_z=\frac{h^2}{\pi}\frac{\rho c_V}{\kappa_\perp}, \qquad \tau_r=\frac{a^2}{\mu^2}\frac{\rho c_v}{\kappa_\parallel}.
\end{equation}
The factor $\mu^2$ is related to the geometry and is taken to be 10, corresponding to a circular geometry \cite{baglioni2023}. If min$(\tau_z,\tau_r)=\tau_{\text{mech}}$ the dissipation will be the highest because temperature variations happen at the same timescale as the strain variations, such that thermal expansion forces efficiently contribute to damping. However, if $\tau_z,\tau_r\gg\tau_{\text{mech}}$, the heat gradients relax without dissipating energy. Similarly, if $\tau_z$, $\tau_r\ll\tau_{\text{mech}}$ the heat gradients do not have time to equilibrate, there is no heat flow, such that no time-dependent thermal damping forces occur and dissipation is low. Finally, if $\tau_r=\tau_{\text{mech}}$, but $\tau_z\gg\tau_{\text{mech}}$, or vice versa, the thermal gradients relax quickly via the fastest route without dissipating energy. We stress that we only qualitatively discuss the relaxation times to provide insight and an intuitive picture, all calculations of the dissipation in this work come from Eqs. (\ref{eq:deltaEsum}-\ref{eq:dissipation}).
\section{Thermal properties with magnetic phase transition}
\label{sec:thermal}
So far we have only considered the elastic and thermal degrees of freedom, in this section we include the magnetic ones. To quantitatively explain the thermal damping of the system around the phase transition, we develop a scheme of merging the magnetic contribution into the thermoelastic dynamics by generalizing the traditional Gr\"uneisen relation to include the magnetoelastic interaction \cite{Pulvirenti1996, Gomes2019}. Equipped with this tool we predict the temperature dependence for the linear expansion coefficient over a wide range of temperatures across the magnetic phase transition. 

Before focusing on the effect of magnetoelasticity, we first review the thermal and elastic properties of FePS$_3$. For that, we take experimental measurements from the literature and match them with a simple model accounting for the contributions of phonons, magnons, and the phase transition shown in Appendix \ref{appendix:thermal_props}.

\begin{figure}[h]
\includegraphics[width=0.4\textwidth]{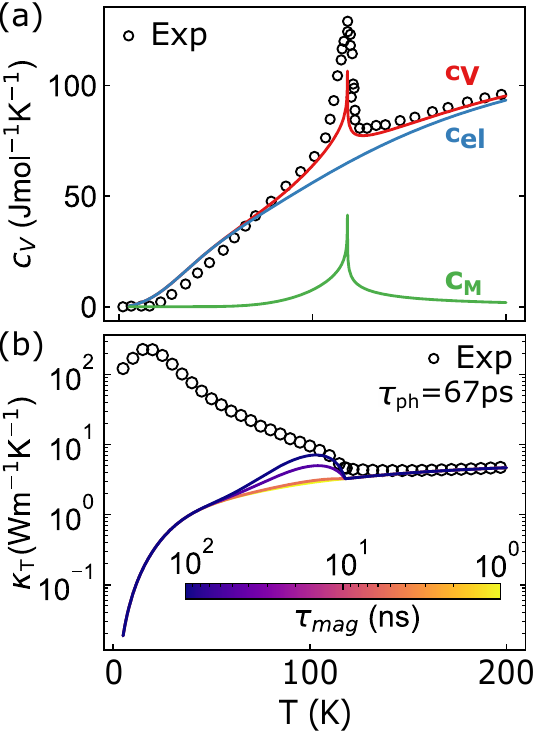}
\caption{Thermodynamic properties of FePS$_3$.
    \textbf{a)} Specific heat, $c_V$, computed as a sum of the elastic ($c_{\text{el}}$) and magnetic ($c_{\text{M}}$) contributions, respectively in blue and green. It is compared with experimental measurements from \cite{Takano2004} (black circles).
    \textbf{b)} Thermal conductivity as a function of temperature for several magnon lifetimes and phonon lifetime $\tau_{\text{ph}}$=67ps, see App. \ref{appendix:thermal_props} Eqs. (\ref{eq:kappa_D})-(\ref{eq:kappa_m}). It is compared with experimental data in bulk from \cite{Haglund2019} (black circles).}
\label{fig:fig3}
\end{figure}

The material parameters are shown in Table \ref{tab:elastic-thermal} and \ref{tab:magnetic}, and the fitting parameters in Table \ref{tab:fit}. In Fig. \ref{fig:fig3} a) we show, in red, the calculated specific heat that comes from the elastic component in blue (App. \ref{appendix:thermal_props}, Eq. (\ref{eq:cv_elastic})) and magnetic components in green (App. \ref{appendix:thermal_props}, Eqs. (\ref{eq:c_magnon})-(\ref{eq:c_ising})), which is compared to experimental data in black \cite{Takano2004}. The magnetic contribution has the magnon and Ising contributions, which show an anomaly at the magnetic phase transition temperature, $T_N=118 K$. The results, with no fit parameters, are in good agreement with experimental measurements in the bulk \cite{Takano2004}. The thermal conductivity results are shown in panel b), showing disagreement between the model and experiments in the low-temperature regime \cite{Haglund2019}. We attribute this to the assumption of a temperature-independent lifetime of the quasiparticles. The best agreement is found for a phonon lifetime of 67 ps and a magnon lifetime of 100 ns, which corresponds to the dark blue curve. According to the literature, the value for the phonon lifetime is reasonable, but the one for magnons is several orders of magnitude higher than expected \cite{xufei2018}, this is probably due to the unrealistic assumption of temperature-independent lifetimes. 
\begin{table}[h]
\centering
\renewcommand{\arraystretch}{1.4} 
\caption{Elastic and thermodynamic parameters of bulk FePS$_3$ \cite{Takano2004, siskins2020}. See App. \ref{appendix:thermal_props} for more information.}
\label{tab:elastic-thermal}
\begin{tabular}{cccccccc}
\hline
~~~$\rho$~~~     & ~~~E~~~   & ~~~$\nu$~~~ & ~~~$\gamma_\text{E}$~~~ & ~~~$\beta_\text{T}$~~~ & ~~~$\Bar{v}$~~~ & ~~~$\theta_\text{D}$~~~   \\ \hline
3375  & 103 & 0.304 & 1.798  & $1.14\cdot10^{-11}$ & 3823 & 236 \\
kg$\cdot$m$^{-3}$ & GPa &    &    & Pa$^{-1}$  & m$\cdot$s$^{-1}$ & K  \\ \hline
\end{tabular}%
\end{table}
\begin{table}[h]
\centering
\renewcommand{\arraystretch}{1.3} 
\parbox{.4\linewidth}{
\centering
\caption{Magnetic parameters of FePS$_3$ \cite{Wyzula2022}. See App. \ref{appendix:thermal_props} for more information.}
\label{tab:magnetic}
\begin{tabular}{cccc}
\hline
J & $T_\text{N}$ & g    & $\gamma$          \\ \hline
2 & 118 K & 2.15 & g$\gamma_e/2$ \\ \hline
\end{tabular}%
}
\hspace{0.2in}
\parbox{.45\linewidth}{
\centering
\caption{Extracted magnetic parameters. See App. \ref{appendix:thermal_props} for more information.}
\label{tab:fit}
\begin{tabular}{ccc}
\hline
$\gamma_\text{M}$  & $H_\text{E}$      & $H_\text{A}$      \\ \hline
$4\gamma_\text{E}$ & $69 \mu_0$ & $138\mu_0$ \\ \hline
\end{tabular}
}
\end{table}

Now we turn our attention to calculating the thermal expansion coefficient, where we account for the effects of magnetostriction. We include the magnetoelastic coupling in the total free energy and analyze the hybrid system with magnetic, elastic, and thermal dynamics \cite{Argyle1967}. The magnetoelastic coupling energy in general has the form \cite{Shapira1976, Shapira1978}
\begin{equation}
    F_{\text{MEC}}=-N_s n_c\frac{\partial J}{\partial V}\braket{\boldsymbol{S}_i\cdot\boldsymbol{S}_j}\epsilon,
\end{equation}
where $N_s$ is the number of spins per unit volume, $n_c$ is the coordination number, $J$ the exchange constant and $\epsilon$ is the volumetric strain. The two-spin correlation function $\braket{\boldsymbol{S}_i\cdot\boldsymbol{S}_j}$ indicates the average over space and time for any two nearest neighboring spins. By adding this contribution to the free energy, we express the total free energy, considering both the thermal and magnetoelastic coupling,
\begin{equation}
\begin{split}
    F &=  F_0 - K_T\alpha_E \Theta \epsilon + \frac{1}{2}K_T(\epsilon)^2\\
    &+G\sum_{ij}\left(\epsilon_{ij}-\frac{1}{3}\epsilon\delta_{ij}\right)^2 -N_s n_c\frac{\partial J}{\partial V}\braket{\boldsymbol{S}_i\cdot\boldsymbol{S}_j}\epsilon,
\end{split}
\end{equation}
where $\epsilon_{ij}$ is the strain tensor, $G$ is the shear modulus and $K_T$ the bulk modulus. The strain in equilibrium can be derived from $\partial F/\partial\epsilon=0$, for a material with isotropic elastic properties, leading to the combined effect on volume changes due to both thermal expansion and magnetostriction
\begin{equation}
    \epsilon=\alpha_E\Theta-\beta_T N_s n_c\frac{J}{V}\gamma_M\braket{\boldsymbol{S}_i\cdot\boldsymbol{S}_j},
\end{equation}
where $\beta_T$ is the compressibility ($1/K_T$) and $\gamma_M$ is the magnetic Grüneisen constant describing the volume dependence on the exchange coupling strength. It can be written in the form
\begin{equation}
    \gamma_M=-\frac{V}{J}\frac{\partial J}{\partial V}.
\end{equation}
The part of volume changes due to the magnetostriction is proportional to the two-spin correlation function which can be changed by the variation of either temperature or external field. The temperature increase leads to the decay of spin correlation and results in magnetostriction expansion. Since the magnetic energy derived from the Heisenberg Hamiltonian $H=-2J\sum\boldsymbol{S}_i\cdot\boldsymbol{S}_j$ that is $E_M=-N_s n_cJ\braket{\boldsymbol{S}_i\cdot\boldsymbol{S}_j}$, it is reasonable to define the magnetic specific heat as \cite{Argyle1967}
\begin{equation}
    c_M = -\frac{N_s}{V} zJ\frac{\partial \braket{\boldsymbol{S}_i\cdot\boldsymbol{S}_j}}{\partial T}.
\end{equation}
As a result, the deviation of local spin coherence due to the small change of local temperature, $\Theta$, is $c_M\Theta=-N_s zJ\braket{\boldsymbol{S}_i\cdot\boldsymbol{S}_j}$ and the total volume change can be succinctly expressed in the form
\begin{equation}
    \epsilon =\alpha_E\Theta + \beta_T\gamma_M c_M\Theta = (\alpha_E+\alpha_M)\Theta\equiv \alpha_T\Theta,
\end{equation}
taking into account that $\Theta$ is small enough that $\alpha_E$ and $\alpha_M$ do not change from $T$ to $T_0$. By merging the magnetoelastic coupling into the free energy, we find how the magnetic contribution gets included in the elastic thermal expansion coefficient $\alpha_T=\alpha_E+\alpha_M$.

\begin{figure}[h]
\includegraphics[width=0.4\textwidth]{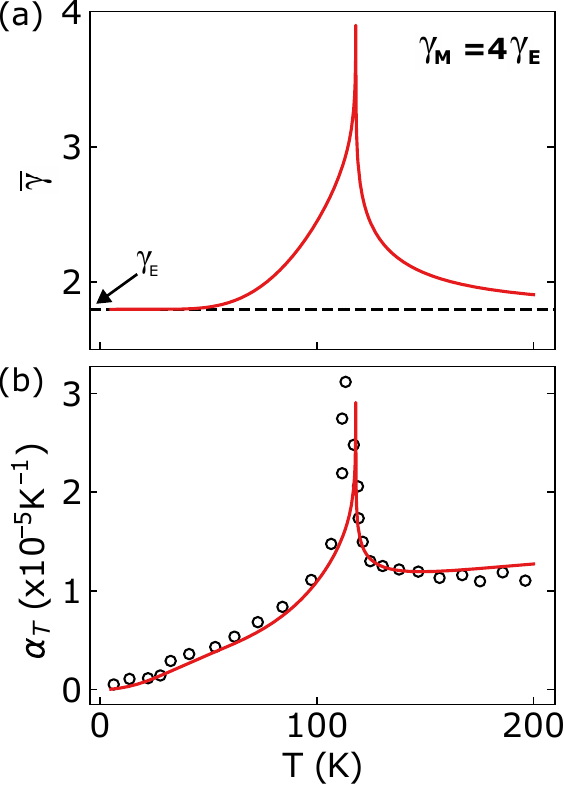}
\caption{
\textbf{a)} Temperature dependence for the effective Grüneisen parameter $\overline{\gamma}$ derived from Eq. (\ref{eq:eff_gamma}) and using values from Tables \ref{tab:elastic-thermal}-\ref{tab:fit}. The elastic Grüneisen parameter is $\gamma_E$ = 1.798 and the magnetic parameter $\gamma_M=4\gamma_E$, fitting the experimental data shown in \textbf{b)}. The dashed line at $\overline{\gamma}=\gamma_E$ shows that the effective Grüneisen parameter is not affected by the magnetic interaction both in the low and high-temperature regimes. \textbf{b)} Thermal expansion coefficient derived from Eq. (\ref{eq:alpha}), with the magnetic Grüneisen parameter as a fitting parameter compared with experimental data from \cite{siskins2020} (black circles).}
\label{fig:fig4}
\end{figure}

The magnetic Grüneisen relation $\alpha_M=\beta_T\rho\gamma_Mc_M$
is similar to the elastic counterpart, $\alpha_E=\beta_T\gamma_E\rho c_E$, where $c_E$ is the specific heat due to the phonon bath. Therefore the overall thermal expansion coefficient for the hybrid system can be written into the form
\begin{equation}
    \alpha_T=\beta_T\rho(\gamma_E c_E +\gamma_M c_M)=\beta_T\rho\overline{\gamma}c_V
    \label{eq:alpha}
\end{equation}
with $\alpha_T$ and $c_V=c_E+c_M$ the thermal observables which can be measured and predicted based on this theory. The effective Grüneisen parameter, $\overline{\gamma}$, is defined as 
\begin{equation}
    \overline{\gamma}=\frac{\gamma_E c_E+\gamma_M c_M}{c_E + c_M}.
    \label{eq:eff_gamma}
\end{equation}
The elastic Grüneisen parameter can be calculated from the Poisson ratio for isotropic materials: $\gamma_E=1.5(1+\nu)/(2-3\nu)$ \cite{sanditov2011}, and the magnetic Grüneisen parameter can be extracted from fitting the measured thermal expansion coefficient as shown in Fig. \ref{fig:fig4}. The effective Grüneisen parameter has a strong temperature dependence as shown in Fig. \ref{fig:fig4} for FePS$_3$. This strong temperature dependence is not caused by variations in the elastic and magnetic Grüneisen parameters but is mainly a result of the temperature dependence of the magnetic and elastic specific heats. The effective Grüneisen parameter shows an anomaly at the phase transition and settles at the value of the elastic parameter, $\gamma_E$, far from the transition temperature, $T_N$. 

This is of importance for TED as the relaxation strength, $\Delta_E$, is proportional to $\alpha_\text{T}^2/c_V$ such that 
\begin{equation}
    \Delta_E\propto \frac{\alpha^2_{\text{T}}}{c_V}=\frac{(\beta_T\rho \bar{\gamma}c_V)^2}{c_V}=(\beta_{\text{T}}\rho)^2\bar{\gamma}^2c_V.
\end{equation}
As we have seen, both $\bar{\gamma}$ and $c_V$ have strong temperature dependences, which will be reflected in $\mathcal{Q}^{-1}$.

\section{Results}
\label{sec:results}
\begin{figure*}[htb]
    \includegraphics[width=0.9\textwidth]{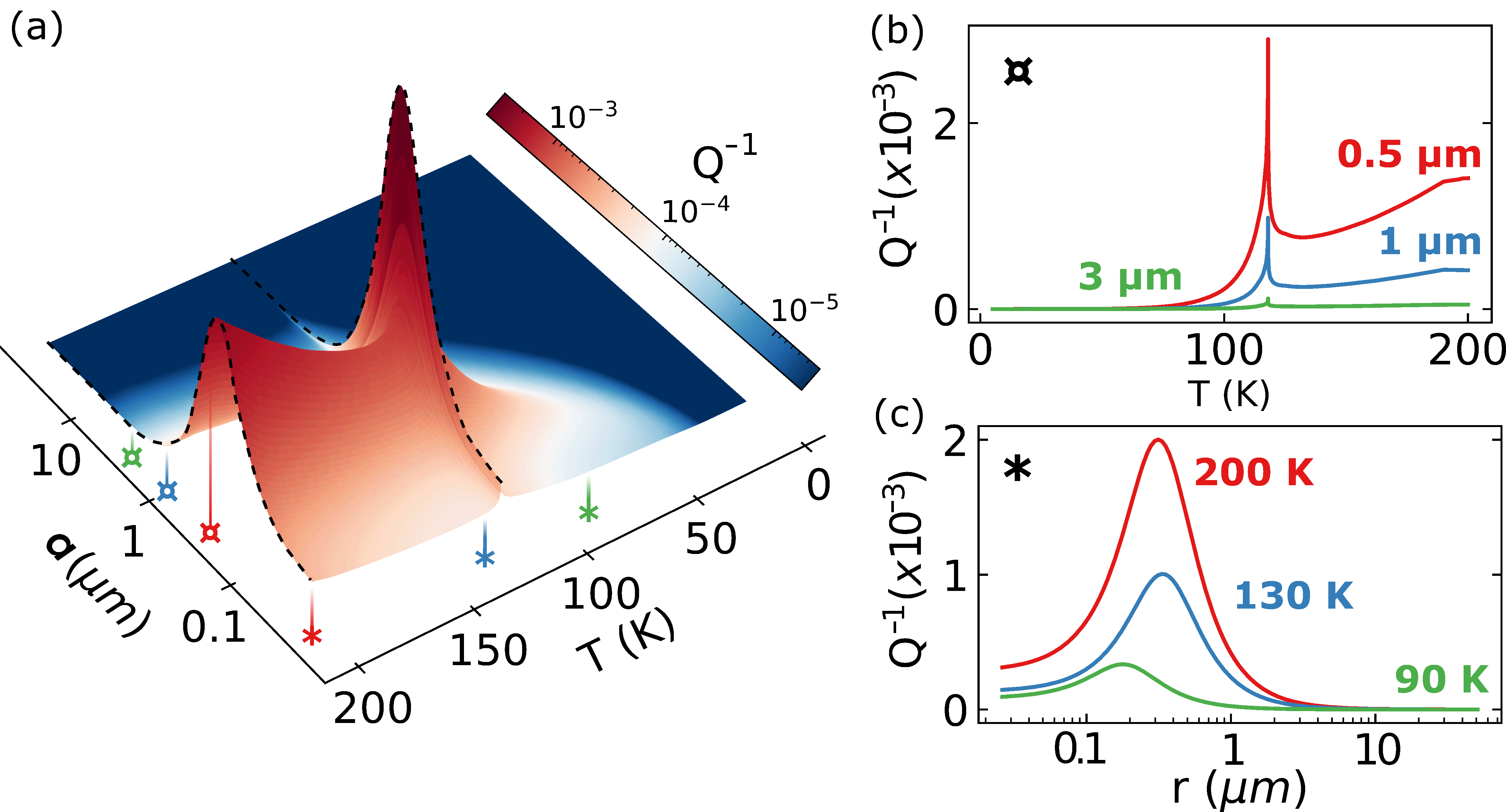}
    \caption{$\mathcal{Q}^{-1}_{\text{TED}}$ for the Z and RZ models, radius dependence. \textbf{a)} Temperature and radius dependence of the predicted dissipation for a 45 nm thick flake. \textbf{b)} Cross sections of the surface in a) at radii 0.5, 1, and 3 $\mu$m indicated, respectively, by red, blue, and green \textcurrency. \textbf{c)} Cross sections of the surface in a) at temperatures of 90, 130, and 200 K indicated, respectively, by green, blue, and red $\ast$. The main features are the peak at the Néel temperature in the constant radius traces and the Debye peaks in the constant temperature traces, where the thermal time constant matches the resonance frequency of the plate.}
    \label{fig:fig5}
\end{figure*}
Having computed all the thermal properties we can now evaluate the dissipation from Eqs. (\ref{eq:deltaEsum})-(\ref{eq:dissipation}), where the temperature dependence comes into play through the temperature-dependent thermodynamic properties discussed previously. Fig. \ref{fig:fig5} a) shows the temperature and radius dependence of the inverse quality factor. It is plotted as a surface on a linear scale and color-coded on a logarithmic scale. The thickness of the flake is $45\mathrm{nm}$, the specific heat and thermal expansion coefficient are those calculated and plotted in Fig. \ref{fig:fig3} and Fig. \ref{fig:fig4}, and the temperature dependence of the thermal conductivity is the one measured in bulk by A. Halmund~\cite{Haglund2019}, scaled to fit the in-plane and through-plane values measured by \cite{Kargar2020}. The scaling is done such that $\kappa_i(T)=\kappa_{\text{bulk}}(T)\frac{\kappa_i(RT)}{\kappa_{\text{bulk}}(RT)}$, with $i=r,z$ and $RT$ room temperature. In this way, despite the lack of experimental measurements of the T-dependent $\kappa_\parallel$ and $\kappa_\perp$,  we can include the anisotropy, under the assumption that the scattering mechanisms evolve with temperature in the same way for both directions.

In panel b), we focus on three temperature-dependent traces with radii of 0.5, 1, and 3 $\mu$m. All three traces show the same feature: a peak in dissipation at the phase transition. This peak coincides with the anomaly at the phase transition of the thermal expansion coefficient, which appears squared in the strength of the dissipated energy in Eq. (\ref{eq:deltaEsum}). On top of that, there is another effect due to the interplay of the thermal time constant and the mechanical resonant frequency explained previously in Section \ref{sec:model}. The out-of-plane thermal relaxation time $\tau_z$ also shows an anomaly at the phase transition, which brings it closer to $\tau_{\text{mech}}$, thus closer to resonance, enhancing the dissipation. A qualitative understanding of this effect can be obtained by looking at some typical numbers. Due to the small thickness of the van der Waals materials used in nanomechanical resonators (tens of nanometers), $\tau_z$ is very small, of the order of tens of picoseconds at low temperatures and reaching hundreds of picoseconds at the phase transition and higher temperatures. This is quite far from $\tau_{\text{mech}}$ which is hundreds of nanoseconds for a 5 $\mu$m drum. Even if the resonance condition is not met, at the phase transition the time constants are the closest and thus the dissipation is the most efficient, resulting in the peak at the Néel temperature of FePS$_3$. 

Panel c)  shows the radius dependence of three traces at constant temperatures of 200, 130, and 90 K. They show the characteristic Lorentzian shape of a Debye peak, widely discussed in literature \cite{zener1937, Lifshitz2000, Schmid2016}. As the radius of the plate changes, its resonance frequency scales as $1/a^2$ in the absence of tension, such that at some point it becomes resonant because $\tau_z$ is radius independent. This peak in dissipation behavior is what we see in each of the traces of panel c). For each of the temperatures, $\tau_z$ has a different value, and thus, the peak in dissipation condition is met at different radii. 

\begin{figure}[htb]
    \includegraphics[width=0.48\textwidth]{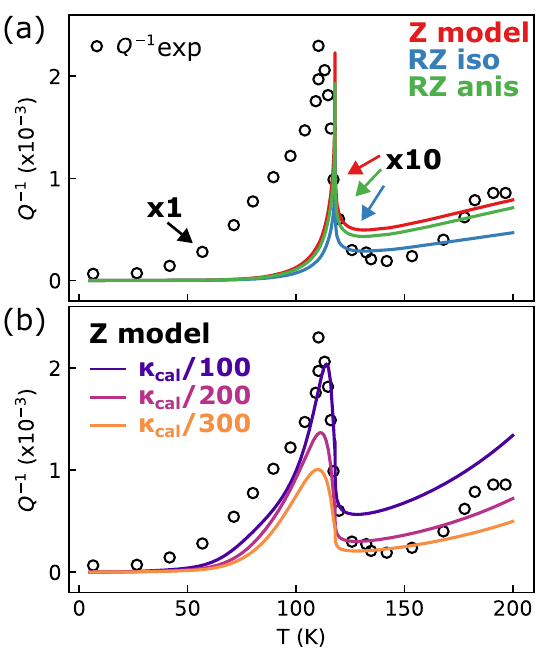}
    \caption{\textbf{a)} Predictions of the dissipation according to the Z and RZ models with the experimental values of bulk thermal conductivity. They are compared with experimental data from \cite{siskins2020}. None of the models reproduce quantitatively the experimental data. Qualitatively, both models show a peak at the Néel temperature, but none predict the rise of the dissipation above 150 K seen in the experiment. \textbf{b)} Comparison of the Z model according to the calculated thermal conductivity divided by factors of 100, 200, and 300.}
    \label{fig:fig6}
\end{figure}

We can also compare the predicted temperature-dependent dissipation with previously reported measurements in this system \cite{siskins2020}. In Fig. \ref{fig:fig6} a) we show the calculations according to the Z model, the RZ model with an isotropic thermal conductivity corresponding to the bulk measured value, and that adapted to take into account the reported anisotropy in FePS$_3$. All the models predict a similar temperature dependence with a sharp peak at the phase transition and a slightly increasing tendency above the Néel temperature. If we compare it with the experimental data shown with black dots, we notice two things: 1) the peak in the experimental data is much broader, with higher dissipation in the antiferromagnetic state, and 2) the models predict values for the dissipation at, and above the phase transition, 10 times smaller than the measured values. This is depicted by scaling the modeled curves by a factor of 10. 

If instead we take into account a thermal conductivity modeled with a constant lifetime for both phonons and magnons and include a proportionality factor that reduces its value, the magnitude of the dissipation changes appreciably as presented in panel b). The calculated dissipation curves correspond to thermal conductivities reduced by factors of 100, 200, and 300, respectively. These curves present a better agreement with the experimental data, both in shape and magnitude. 

\begin{figure}[h]
    \includegraphics[width=0.48\textwidth]{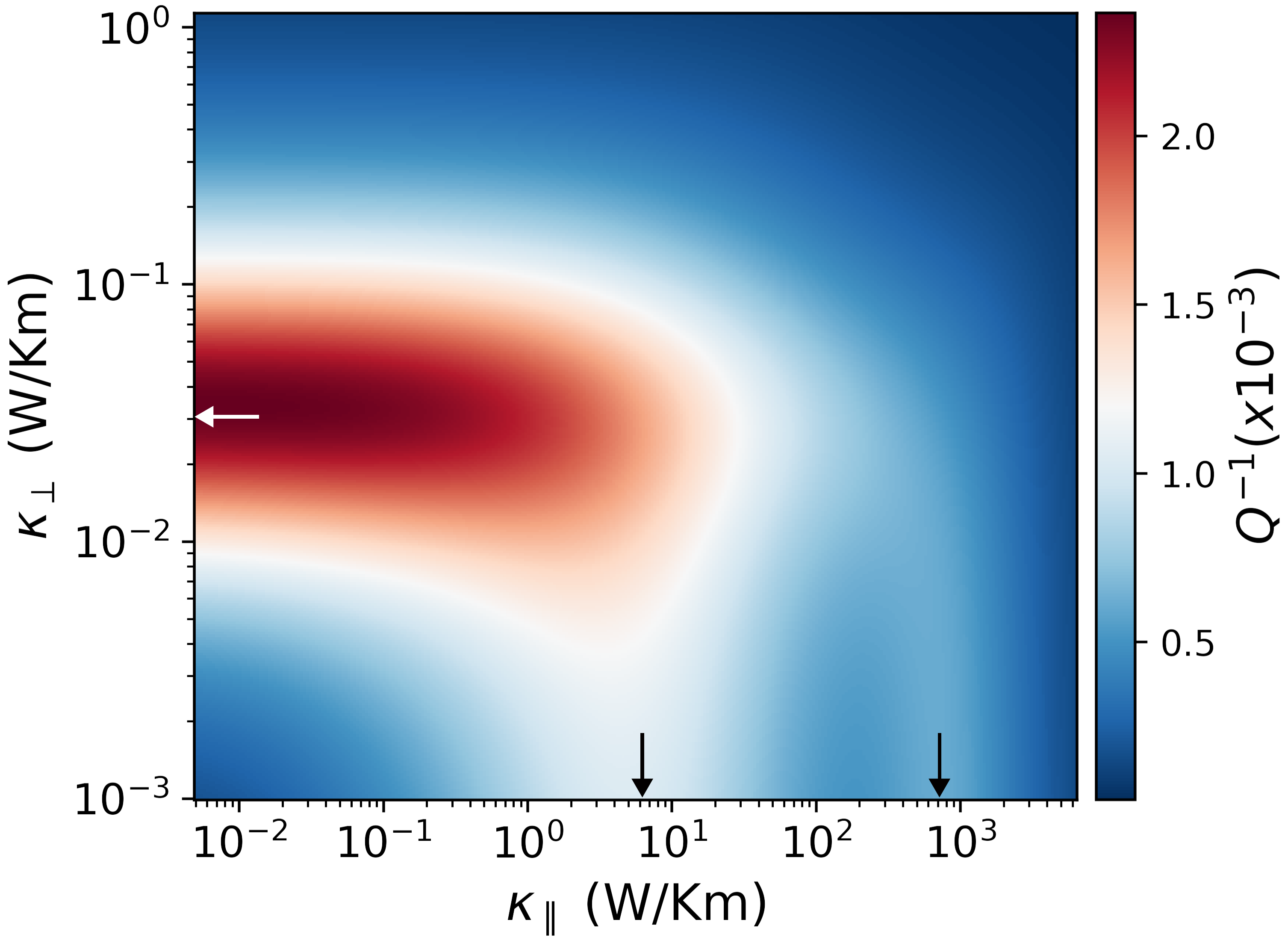}
    \caption{Dissipation as a function of in-plane and out-of-plane thermal conductivities. The drum geometry is 5$\mu$m radius and 40 nm thickness and the thermal properties are taken at T=200 K. The white and black arrows point at values of $\kappa_\perp$ or $\kappa_\parallel$ where the dissipation peaks due to a match of resonance frequency and thermal relaxation time.}
    \label{fig:fig7}
\end{figure} 

Finally, we analyze the consequences of varying the in-plane versus out-of-plane thermal conductivities for TED. We expect that varying $\kappa_\perp$ and $\kappa_\parallel$ will only modify $\tau_z$ and $\tau_r$ respectively, leaving $\tau_\text{{mech}}$ unchanged. We will then see all the situations described at the end of Section \ref{sec:model}, namely the different resonances between thermal relaxation rates and the mechanical mode, and regions where heat gradients relax without dissipating energy. Fig. \ref{fig:fig7} shows the calculations for the dissipation of a drum with a thickness of 40nm and 5$\mu m$ radius at a fixed temperature of 200 K. This drum has $f=15.9$ MHz or $\tau_{\text{mech}}=63$ ns. For values of $\tau_r\ll\tau_{\text{mech}}$ (low $\kappa_\parallel$), we can observe the peak dissipation condition when $\kappa_\perp$ is such that $\tau_z=\tau_{\text{mech}}$ indicated with a white arrow. Similarly, for values of $\tau_z\ll\tau_{\text{mech}}$ (low $\kappa_\perp$), we can observe the peak dissipation condition when $\kappa_\parallel$ is such that $\tau_r=\tau_{\text{mech}}$ indicated with black arrows. In this case, we notice that there are two peak dissipation conditions, which cannot be explained by what has been discussed so far. This fact hints that there are two thermal relaxation times in the $r$-direction. When $\tau_z\gg\tau_{\text{mech}}$ or $\tau_r\gg\tau_{\text{mech}}$ (high $\kappa_\perp$ or $\kappa_\parallel$), heat relaxes quickly and there is little dissipation. Contrarily, when $\tau_z,\tau_r\ll\tau_{\text{mech}}$ (low $\kappa_\perp$ and $\kappa_\parallel$) heat does not have time to relax and the dissipation is also low.

A possible hypothesis for having two in-plane thermal relaxation times is related to the fact that the strain generates heat in localized regions of the drum, namely, at the center and the edges (Fig. \ref{fig:fig2}). The time it takes for these different localized heated regions to relax into the substrate is different, as they travel different lengths through the plate. We can thus reformulate the in-plane thermal time constant as $\tau_{\parallel}^*=\dfrac{(\lambda a)^2}{\mu^2}\dfrac{\rho c_v}{\kappa_{\parallel}}$, where $\lambda$ is a parameter that renormalizes the distance traveled by the heat. If $\lambda=1$ it describes heat traveling from the center of the plate, and if we set $\lambda\simeq 0.1$, it corresponds to the heat generated in the edges traveling a shorter length of the drum. In this last case, we get a thermal time constant 100 times faster. Thus, these two constants will match $\tau_{\text{mech}}$ for different values of $\kappa_\parallel$, 2 orders of magnitude apart from each other. That is the distance we observe between the black arrows in Fig.\ref{fig:fig7}. A way to validate this hypothesis could be computing the dissipation following the same procedure as in this work but for higher order modes, such that for every extra antinode in the strain we should see an extra peak in the dissipation as a function of $\kappa_\parallel$.

\section{Discussion and Outlook}
\label{sec:discussion}

In this work, we developed a new model for thermoelastic damping that incorporates anisotropic thermal conduction, specifically suitable for van der Waals materials. This model improves upon traditional Zener and Lifshitz-Roukes models by accounting for the anisotropic heat flow typical in materials with high in-plane to out-of-plane conductivity ratios. We focused on the fundamental vibrational mode of nanomechanical resonators made of FePS$_3$, using classical thermodynamics to compute key thermal properties like specific heat, thermal conductivity, and thermal expansion coefficient. Incorporating the effective Grüneisen parameter allowed us to capture the effects of magnetostriction, crucial for understanding dissipation near the magnetic phase transition. Our results reveal temperature- and size-dependent dissipation behavior, showing both the Debye peak and an anomaly at the phase transition, which aligns well with experimental observations. However, quantitative agreement with experimental data was only achieved when the out-of-plane thermal conductivities were reduced by two orders of magnitude compared to bulk measurements, indicating a need for better characterization of thermal conductivity in suspended resonators.

The TED model we propose is based on diffusive phonon and magnon propagation and remains valid as long as the device dimensions exceed the mean free path of heat-carrying quasiparticles. At 100 K, the mean free path for phonons is about 4.5 nm, well below the minimum drum size analyzed in this work, ensuring the model's applicability. While we relied on bulk measurements or theoretical predictions for thermal conductivity, future studies could benefit from direct, temperature-dependent measurements in nanoscale vdW devices, especially for in-plane and out-of-plane thermal conductivities. Extending the current analysis to higher-order vibrational modes would allow for a more comprehensive exploration of TED, potentially uncovering additional thermal relaxation times.
    
\begin{acknowledgments}
This publication is part of the project ``Ronde Open Competitie XL" (file number OCENW.XL21.XL21.058) which is financed by the Dutch Research Council (NWO). X. Z. thanks for the support from the China Postdoctoral Science Foundation, Grant No. 2022M710402. M.\v{S}. acknowledges funding from the Ministry of Education, Singapore, under its Research Centre of Excellence award to the Institute for Functional Intelligent Materials, Project No. EDUNC-33-18-279-v12.
\end{acknowledgments}

\bibliographystyle{apsrev4-2}
\bibliography{main}

\appendix

\section{Dissipation Mechanisms}
\label{appendix:dissipation}
In Table~\ref{tab:dissipation} we show a list of possible dissipation mechanisms that could be relevant in a magnetic nano-mechanical resonator. This list was elaborated to pinpoint the most relevant mechanisms participating in our system of study. 
\begin{table*}[ht]
    \caption{Summary of dissipation mechanisms for nanomechanical resonators with magnetic insulating membranes. For each mechanism, we include a description, the presence of temperature dependence, a comment on the relevance, and references.}
    \label{tab:dissipation}
    \resizebox{0.95\textwidth}{!}{%
    \begin{tabular}{ccccc}
        \hline
        \textbf{\begin{tabular}[c]{@{}c@{}}Dissipation \\ mechanism\end{tabular}} &
        \textbf{Description} &
        \textbf{T dep} &
        \textbf{Relevance} &
        \textbf{References} \\ \hline
    \begin{tabular}[c]{@{}c@{}}Medium \\ losses\end{tabular} &
    \begin{tabular}[c]{@{}c@{}}Interaction with surrounding \\ liquid or gas.\\ $~$\end{tabular} & Y &
    \begin{tabular}[c]{@{}c@{}}Possible interaction with \\ trapped gasses.\\ $~$\end{tabular} & \cite{Steeneken2021} \\
    \begin{tabular}[c]{@{}c@{}}Ohmic \\ loss\end{tabular} &
    \begin{tabular}[c]{@{}c@{}}Interactions between free electrons in\\  the membrane and charges in \\ the substrate or gate electrode.\\ $~$\end{tabular} & Y &
    \begin{tabular}[c]{@{}c@{}}Improbable due\\ to low conductance\\ of the used materials.\end{tabular} & \cite{seoanez2007} \\
    \begin{tabular}[c]{@{}c@{}}Acoustic \\ radiation\end{tabular} &
    \begin{tabular}[c]{@{}c@{}}Transfer of mechanical energy \\ to the anchoring substrate.\\ $~$\end{tabular} & N &
    \begin{tabular}[c]{@{}c@{}}It is not expected\\ to participate in any \\ relevant features.\\ $~$\end{tabular} & \cite{ Bachtold2022,Steeneken2021} \\
    \begin{tabular}[c]{@{}c@{}}Adhesion \\ loss\end{tabular} &
    \begin{tabular}[c]{@{}c@{}}Loss at the edges by formation \\ and destruction of bonds between \\ the membrane and the substrate.\\ $~$\end{tabular} & N &
    \begin{tabular}[c]{@{}c@{}}Unknown. Out of the \\ scope of this project.\\ Low for graphene.$~$\end{tabular} & \cite{seoanez2007} \\
    \begin{tabular}[c]{@{}c@{}}Interlayer \\ friction\end{tabular} &
    \begin{tabular}[c]{@{}c@{}}Friction between layers  \\ due to weak interlayer \\ vdW interaction.\\ $~$\end{tabular} & Y &
    \begin{tabular}[c]{@{}c@{}}Unknown. Out of the \\ scope of this project.\\ Low for graphene.\\$~$\end{tabular} & \cite{Will2017,Wei2022} \\ Defects &
    \begin{tabular}[c]{@{}c@{}}Modelled as two level systems, \\ defects can form structures that \\ can be excited and store energy.\\ $~$\end{tabular} & Y &
    \begin{tabular}[c]{@{}c@{}}Improbable. It is only \\ predominant for \\ amorphous insulators.\\ $~$\end{tabular} & \cite{Bachtold2022, seoanez2007, Mohanty2002} \\
    \begin{tabular}[c]{@{}c@{}}\textbf{Thermoelastic} \\ \textbf{damping}\end{tabular} &
    \begin{tabular}[c]{@{}c@{}}Local temperature gradients due \\ to compression and expansion \\ lead to irreversible heat flow.\\ $~$\end{tabular} & Y &
    \begin{tabular}[c]{@{}c@{}}Important.\\ Observed in\\ references.\end{tabular} & \cite{siskins2020, zener1937, Lifshitz2000} \\
    \begin{tabular}[c]{@{}c@{}}Magnetoelastic \\ damping\end{tabular} &
    \begin{tabular}[c]{@{}c@{}}Magnons become a heat \\ reservoir via interaction of \\ phonons and spins.\\ $~$\end{tabular} & Y &
    \begin{tabular}[c]{@{}c@{}}Unknown. \end{tabular} & \cite{Shklovskij2021, Wang2020}\\
    \begin{tabular}[c]{@{}c@{}}Phonon-phonon \\ coupling (Akhiezer)\end{tabular} &
    \begin{tabular}[c]{@{}c@{}}Coupling of the resonant \\ mode and the phonon bath, or \\ with higher order modes.\\ $~$\end{tabular} & Y &
    \begin{tabular}[c]{@{}c@{}}Unknown. Out of the \\ scope of this project. \end{tabular} & \cite{Bachtold2022} \\
    \begin{tabular}[c]{@{}c@{}}Photon-phonon \\ coupling\end{tabular} &
    \begin{tabular}[c]{@{}c@{}}It allows energy transfer from\\ the membrane to optical modes\\ (cavity or free-space).\\ $~$\end{tabular} & Y &
    \begin{tabular}[c]{@{}c@{}}Improbable due to \\ low coupling and \\ off-resonance.\end{tabular} & \cite{Aspelmeyer2014} \\
    \begin{tabular}[c]{@{}c@{}}Many-body \\ electron phenomena\end{tabular} &
    \begin{tabular}[c]{@{}c@{}}Kondo effect, quantum Hall effect, \\ Fabry-Perot interference, and \\ Aharonov-Bohm oscillations. \\ $~$\end{tabular} & Y &
    \begin{tabular}[c]{@{}c@{}}None. Our devices\\ are not in such\\ quantum regimes.\end{tabular} & \cite{Bachtold2022} \\
    \begin{tabular}[c]{@{}c@{}}Casimir \\ force\end{tabular} &
    \begin{tabular}[c]{@{}c@{}}It opens a channel for \\ energy transmission from \\ membrane to substrate. \\ $~$\end{tabular} & Y &
    \begin{tabular}[c]{@{}c@{}}None due to the lack \\ of conductivity and \\ geometry factors.\\ $~$\end{tabular} & \cite{Bachtold2022} \\ \hline \end{tabular}%
    }
\end{table*}

\section{Temperature dependent resonant frequency}
\label{appendix:temperature}
As the temperature of the system is varied, tension is exerted on the suspended material which shifts the resonant frequency. The change in resonance frequency has a direct effect on the dissipation and is thus a key component for its calculation. 

We saw in Eq. (\ref{eq:tension}) how to calculate the temperature-dependent tension due to the mismatch in the thermal expansion coefficient of the suspended material and the substrate. The total tension is then $N_{\text{tot}}(T)=N_0+N(T)$, with $N_0$ a uniform constant pretension that we keep as a fitting parameter when comparing to experimental results. Knowing the tension and material parameters, we only need to compute the parameters $\delta_+$ and $\delta_-$ to obtain the temperature-dependent resonant frequency. To obtain these parameters, we simultaneously solve the following equations, \cite{Ma2020}
\begin{equation}
    \delta_-\frac{J_1(\delta_-)}{J_0(\delta_-)} + \delta_+\frac{I_1(\delta_+)}{I_0(\delta_+)} = 0, \quad \text{and} \quad \delta_+^2 - \delta_-^2 = \frac{a^2N}{D}.
\label{eq:deltas}
\end{equation}
Using the computed thermal expansion coefficient for FePS$_3$ (Fig. \ref{fig:fig4}), and that of Si from \cite{Middelmann2015}, we calculate the tension shown in Fig. \ref{fig:fig7} a). And following the explained procedure we obtain the resonant frequency, shown in Fig \ref{fig:fig7} b), with fitting parameter $N_0=6.85$ N/m. In this case, the parameter $N_0$ also includes the acquired tension from room temperature to 200 K, in addition to the pretension existing in the suspended material at room temperature.
\begin{figure}[h]
    \includegraphics[width=0.42\textwidth]{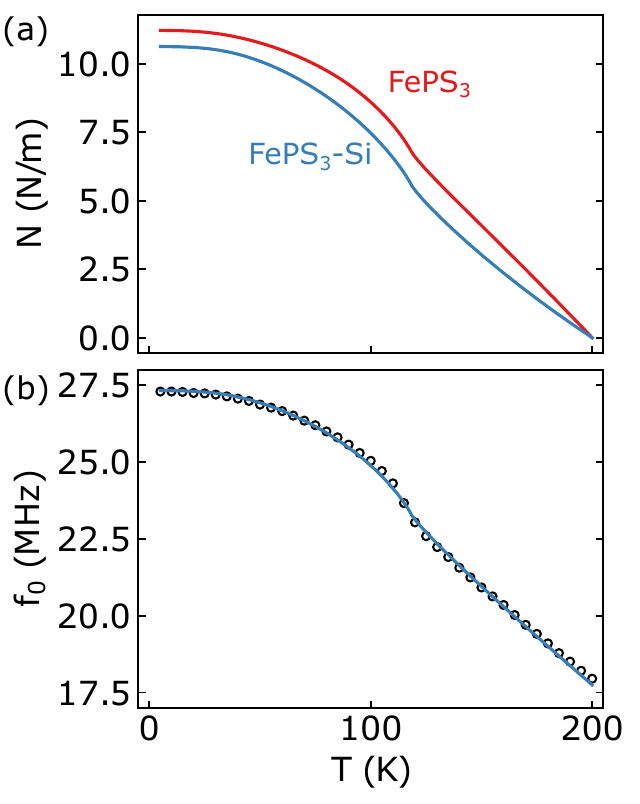}
    \caption{\textbf{a)} Radial tension accumulated in the plate due to thermal expansion. In red the model considers only the expansion of the FePS$_3$ plate, and in blue it corrects for the expansion of the substrate. \textbf{b)} Resonance frequency as a function of the temperature of a 45 nm thick drum fitted according to the plate model with temperature-dependent tension shown in a). The model assumes a constant pretension N$_0$ as a fit parameter. The black dots are experimental data from M. Šiškins et al. (2020) \cite{siskins2020}.}
    \label{fig:fig8}
\end{figure}

\section{Solving the thermoelastic equation}
\label{appendix:solving}
To obtain the temperature field we solve Eq. (\ref{eq:anis_thermoelastic}) by substituting $\epsilon$ from Eq. (\ref{eq:straincomp}) and $\Theta=\Theta_0(r,z)e^{i\omega t}$,
\begin{align}
    \begin{split}
        & \left[\kappa_\perp\frac{\partial^2}{\partial z^2} +  \nabla ^2 + \kappa_\parallel\left(\frac{\partial^2}{\partial r^2} + \frac{1}{r}\frac{\partial}{\partial r}\right)\right] \Theta _0=\\
        & i\omega \left(\rho c_\text{V} \Theta_0 -\beta_{\text{TE}} T_0 z \nabla^2 W\right).
    \end{split}
    \label{eq:2dheat_app}
\end{align}
We assume full thermal contact at the edges: $\Theta_0(r,z)\big|_{r=a}=0$. To find a solution we expand in terms of eigenfunctions of the radial part
\begin{equation}
    \Theta_0(r,z)=\sum_{n=1}^\infty J_0\left(\frac{j_0^n}{a}r\right)Z_n(z),
    \label{eq:eigexp_app}
\end{equation}
with $Z_n$ the unknown solution of the z component. Substituting into Eq. (\ref{eq:2dheat_app}) we get
\begin{equation}
\begin{split}
    &\kappa_\parallel \sum_{n=1}^\infty (\frac{j_0^n}{a})^2 J_0\left(\frac{j_0^n}{a}r\right)Z_n(z) + \kappa_\perp J_0\left(\frac{j_0^n}{a}r\right)Z_n''=\\
    &i\omega\left(\rho c_\text{V}J_0\left(\frac{j_0^n}{a}r\right)Z_n-\beta_{\text{TE}}z\nabla^2W\right).
\end{split}
\end{equation}
Putting all terms with the Bessel function together, and dividing by $\kappa_\perp$, it reads
\begin{align*}
     &\sum_{n=1}^\infty \left[Z_n''-\left(\frac{\kappa_\parallel}{\kappa_\perp}\left(\frac{j_0^n}{a}\right)^2+i\omega\frac{\rho c_\text{V}}{\kappa_\perp} \right)Z_n\right]J_0\left(\frac{j_0^n}{a}r\right) = \numberthis\\
       &-i\omega z\frac{\beta_{\text{TE}}\nabla^2W}{\kappa_\perp}.
\end{align*}
Multiplying both sides by $rJ_0\left(j_0^n r/a\right)$ and integrating over $r$ from $0$ to $a$ we get
\begin{equation}
    Z''_n(z)-b^2_nZ_n(z)=c_n z,
    \label{eq:Zequation}
\end{equation}
with coefficients
\begin{equation}
    \begin{split}
        b_n^2&=\frac{\kappa_\parallel}{\kappa_\perp}\left(\frac{j_0^n}{a}\right)^2+i\omega\frac{\rho c_\text{V}}{\kappa_\perp},\\
        c_n = \frac{-2i\omega}{a^2[J_1(j_0^n)]^2}&\frac{T_0\beta_{\text{TE}}}{\kappa_\perp}\int_0^a\nabla^2 WJ_0\left(\frac{j_0^n}{a}r\right)rdr.
    \end{split}
    \label{eq:coefficients_TE}
\end{equation}
The solution to Eq. (\ref{eq:Zequation}) is
\begin{equation}
    Z_n(z)=\frac{c_n}{b_n^3}\frac{\sinh(b_nz)}{\cosh(b_nh/2)}-\frac{c_n}{b_n^2}z.
\end{equation}
This solution can be readily introduced in Eq. (\ref{eq:eigexp_app}) to obtain the temperature field.

\section{Calculation of the dissipated energy}
\label{appendix:diss_energy}

To express analytically the dissipated energy in terms of material properties and thermodynamical quantities, it is useful to introduce the following variables,

\begin{align*}
    x_n^2=\frac{\kappa_\parallel}{\kappa_\perp}\left(\frac{j_0^n}{a}\right)^2&,\qquad \chi^2=\omega \frac{\rho c_\text{V}}{\kappa_\perp},\\
    b_n^2=(l_n+im_n)^2&,\qquad l_n^2=\frac{x_n+\sqrt{x_n^2+\chi^2}}{2},\quad \numberthis\\
    \text{and}\quad m_n^2&=\frac{-x_n+\sqrt{x_n^2+\chi^2}}{2}.\\
\end{align*}

Taking that into account, the terms in Eq. (\ref{eq:sum_terms_diss}) can be written as
\begin{widetext}
\begin{align*}
    &B_n=x_n\left[\sinh^2(l_nh/2)+\cos^2(m_nh/2)\right],\\
    &C_n=\left(2x_n-\sqrt{x_n^2+\chi^2}\right)\Bigg\{\left(1+\frac{x_n}{\sqrt{x_n^2+\chi^2}}\right)\left[\frac{h}{2}\left(\sinh^2\left(l_n\frac{h}{2}\right)+\cos^2\left(m_n\frac{h}{2}\right)\right)+m_n\sin(m_nh)\right]\\
    &-\frac{x_n}{x_n^2+\chi^2}\sinh(l_nh)\Bigg\}\\
    &+\left(2x_n+\sqrt{x_n^2+\chi^2}\right)\Bigg\{\left(1-\frac{x_n}{\sqrt{x_n^2+\chi^2}}\right)\left[\frac{h}{2}\left(\sinh^2\left(l_n\frac{h}{2}\right)+\cos^2\left(m_n\frac{h}{2}\right)\right)+l_n\sin(m_nh)\right]\numberthis\\
    &+\frac{x_n}{x_n^2+\chi^2}\sin(m_nh)\Bigg\},\\
    &D_n=(x_n^2+\chi^2)^{3/2}\left[\sinh^2(l_nh/2)+\cos^2(m_nh/2)\right]J_1^2(j_0^n),\\
    &I_n=\int_0^a\nabla^2WJ_0(x_nr)rdr=\frac{(\delta-^2+\delta_+^2)(j_0^n)^2J_0(\delta_-)J_1(j_0^n)}{(\delta_-^2-(j_0^n)^2)(\delta_+-(j_0^n)^2)}.
\end{align*} 
\label{eq:factors}
\end{widetext}

\section{Thermal properties with magnetic phase transition}
\label{appendix:thermal_props}
In this Appendix, we analyze in detail the thermal properties of FePS$_3$. We develop a simple scheme to understand the temperature dependence of the specific heat and thermal conductivity. 

In general, below the phase transition, the material’s specific heat, $c_V$, comes from the thermal excitation of the bosons. For ordinary insulators, these are typically phonons and they include magnons for ferromagnetic and antiferromagnetic materials. If the temperature is homogeneous then the Bose-Einstein density of excited bosons is uniformly distributed across the material. However, the existence of a temperature field leads to the excess number of quasi-particles staying out of equilibrium and then transport according to the temperature gradient. If the environment temperature is close to the magnetic phase transition, the coherence of precession between the neighboring spins breaks down and an additional contribution to the specific heat should be considered. The decay of magnetization $M$ as the material heats leads to an accompanying decrease of the effective exchange field $H_E$ and anisotropy field $H_A$ in magnon’s dispersion equation \cite{renzede2019}
\begin{equation}
    \omega_{}=\gamma\mu_0\sqrt{H_A^2+2H_EH_A+H_E^2\sin^2(\pi k/k_m)},
    \label{eq:magnon_dis}
\end{equation}
in which $\gamma=g\mu_B/\hbar$ is the gyromagnetic ratio and $\mu_0$ is the vacuum magnetic constant. The Brillouin zone is limited according to a spherical energy boundary condition, $N=\sum_{\boldsymbol{k}}$, from which $k_m=2\sqrt{\pi}/a$, with $a$ the magnetic lattice constant taken to be 15.94 $\mathring{A}$.  This energy renormalization due to the magnetization decrease should also be incorporated into the calculation of the magnon's specific heat and thermal conductivity \cite{Rezende2014,Shen2018}. Note that the use of Eq. \ref{eq:magnon_dis} is a simplification as we do not account for the specific lattice of FePS$_3$, nor its zigzag magnetic ordering. 

The specific heat due to bosons is given by 
\begin{equation}
    c_\text{V} = \frac{1}{V}\frac{\partial}{\partial T}\sum_{\boldsymbol{k}}\hbar \omega_{\boldsymbol{k}}\Bar{n}_{\boldsymbol{k}}, \quad \Bar{n}_{\boldsymbol{k}}=\frac{1}{e^{\beta \hbar \omega_{\boldsymbol{k}}}-1}.
\end{equation}
where $V$ is the system volume, $\Bar{n}_{\boldsymbol{k}}$ is the Bose-Einstein's equilibrium amount of bosons of energy $\hbar\omega_{\boldsymbol{k}}$ and $\beta=1/k_B T$. The thermal conductivity is defined as the proportionality coefficient for heat flux in response to the temperature gradient, $\boldsymbol{q}=-\kappa\nabla T$. From kinetic transfer theory, this thermal flux can be calculated by \cite{Ashcroft76}
\begin{equation}
    \boldsymbol{q}_\text{T}=-\frac{1}{V}\frac{\partial}{\partial T}\sum_{\boldsymbol{k}}\hbar \omega_{\boldsymbol{k}}\Bar{n}_{\boldsymbol{k}}\tau_{\boldsymbol{k}}(\boldsymbol{\nabla}T\cdot \boldsymbol{v}_{\boldsymbol{k}})\boldsymbol{v}_{\boldsymbol{k}},
    \label{eq:kappa}
\end{equation}
in which an isotropic $\kappa$ can be extracted if the particle velocity $\boldsymbol{v}_k$ is homogeneous to each direction. However, if the velocity of the particles has a directional bias, then $\kappa$ depends on the orientation and thermal transfer shows anisotropy as we encounter in this paper. We do not investigate or model the sources of the anisotropy and focus on the effect it has on the damping.

The elastic specific heat and thermal conductivity can be derived from the low-lying phonon modes with dispersion relation $\omega_{\boldsymbol{k}}=\Bar{v}\boldsymbol{k}$, where $\Bar{v}$ is the Debye averaged velocity, \cite{Pathria2011}
\begin{equation}
    \Bar{v}=\frac{4\pi k_B\theta_\text{D}}{\hbar}\left(\frac{\pi}{6n}\right)^{1/3},
\end{equation}
with $\theta_\text{D}$ the Debye temperature and $n$ the density of atoms. Applying the dispersion relation, a straightforward derivation leads to the Debye specific heat $c_\text{D}$,
\begin{equation}
    c_\text{D}(T)=\frac{\hbar^2}{2\pi}\frac{3}{k_BT^2}\int_0^{k_D}dkkw_k^2\frac{e^{\beta\hbar\omega_k}}{(e^{\beta\hbar\omega_k}-1)^2}.
    \label{eq:cv_elastic}
\end{equation}
Here we assumed a two-dimensional system with still 3 degrees of freedom for the vibrations. The boundary of the Brillouin zone over which it is integrated can be defined by the Debye temperature as $k_\text{D}=k_B\theta_D/\hbar\Bar{v}$. Eq. (\ref{eq:cv_elastic}) provides the specific heat per unit surface, with units in SI J/Km$^2$. However, to compare with experimental measurements it is useful to express it in J/molK for which the specific heat is simply divided by the density of states $n_\text{D}=k_\text{D}^2/4\pi$ and multiplied by the Avogadro constant. In \cite{Takano2004} they suggest the elastic specific heat is best modeled by a mix of Debye and Einstein parts with temperatures $T_D=236$ K and $T_E=523$ K and a combination ratio of 0.54 such that $c_{\text{el}}=0.54c_D+(1-0.54)c_E$ which gives a reasonable estimate for a wide range of temperatures.

From the magnon dispersion relation (\ref{eq:magnon_dis}), the specific heat due to thermal magnons is
\begin{equation}
      c_{\text{mag}}(T)=\frac{\hbar^2}{\pi}\frac{1}{k_BT^2}\int_0^{k_m}dkk\omega_k^2\frac{e^{\beta\hbar\omega_k}}{(e^{\beta\hbar\omega_k}-1)^2},
      \label{eq:c_magnon}
\end{equation}
where two magnon polarizations have been included, due to the antiferromagnetic nature of the magnons. 

To compute the Debye thermal conductivity, $\kappa_\text{D}$, starting from Eq. (\ref{eq:kappa}) and integrating over the available momentum it follows that 
\begin{equation}
    \kappa_\text{D}(T)=\frac{\hbar^2}{4\pi}\frac{3\Bar{v}^2}{k_BT^2}\int_0^{k_\text{D}}dk k \omega_k^2 \tau_k \frac{e^{\beta\hbar\omega_k}}{(e^{\beta\hbar\omega_k}-1)^2}.
    \label{eq:kappa_D}
\end{equation}
The magnon contribution to the thermal conductivity can be extracted from the heat flux equation by taking the velocity of magnons from the dispersion relation, $\boldsymbol{v_k}=\boldsymbol{\nabla_k}\omega_{\boldsymbol{k}}$. It results in
\begin{equation}
\begin{split}
     &\kappa_{\text{mag}}=\left( \frac{\gamma \mu_0 H_E}{8k_m}\right)^2\frac{\pi}{k_BT^2} \\
     &\int_0^{k_m}dk\frac{k\omega_k^2\tau_k\sin^2(k\pi/k_m)}{\sin^2(k\pi/2k_m) + \eta^2 +2\eta}\frac{e^{\beta\hbar\omega_k}}{(e^{\beta\hbar\omega_k}-1)^2}.
\end{split}
\label{eq:kappa_m}
\end{equation}
It is worth mentioning that the specific heat and thermal conductivity as computed here correspond to 2D systems. If one wants to compare the results with measurements of bulk material, the correct quantities are $c_{\text{3D}}=c_{\text{2D}}/h$ and $\kappa_{\text{3D}}=\kappa_{\text{2D}}/h$, with $h$ the thickness of the sample.

As the environment temperature approaches the phase transition regime the magnetic specific heat is dominated by energy absorption for breaking the spin coherence
and due to the nature of second-order phase transition, the anomaly of the specific heat near $T_N$ should be expected \cite{Coey}. The derivation for the anomaly depends on the detailed lattice structure. Since FePS$_3$ is an Ising-type 2D antiferromagnet of the honeycomb (hexagon) lattice~\cite{Lee2016,Wildes2012,Jianmin2021,Lancon2016}, the partition function dues to its magnetic exchange interaction reads~\cite{Houtappel1950,Pathria2011,Matveev1996}
\begin{equation}
\begin{split}
      &\frac{1}{N}\log Z(T)=\log 2~+\\
      &\frac{1}{(4\pi)^2}\!\!\int_0^{2\pi}\!\!\!\!\int_0^{2\pi}\!\!\!d\theta_1 d\theta_2 \log\left(\cosh^3 K\!+\!1\!-\!P_{\boldsymbol{\theta}}\sinh^2 K  \right)\!,
\end{split}
\end{equation}
where $K=2J'/k_BT$ is the normalized temperature, in which $J'=2JS^2$ is the effective exchange coupling. $P_{\boldsymbol{\theta}}=P(\theta_1,\theta_2)=\cos\theta_1 + \cos\theta_2 + \cos(\theta_1+\theta_2)$ is the integrand parameter. The critical point is reached when $\sinh K=\sqrt{3}$, from where the critical temperature is
\begin{equation}
    T_N = \frac{2J'}{k_B\log(2+\sqrt{3})}.
\end{equation}
One can evaluate the effective coupling energy $J'$ based on the measured Neel temperature. From the partition function, the Ising specific heat is readily calculated
\begin{equation}
    c_{\text{Is}}=-\dfrac{d}{dT}\left(\dfrac{d}{d\beta}Z(T)\right), 
\end{equation}
such that
\begin{equation}
\begin{split}
    &c_{\text{Is}}(T)=\frac{k_B K^2}{(4\pi)^2}\int_0^{2\pi}d\theta_1\int_0^{2\pi}d\theta_2 \Bigg\{\\
    &\frac{6\sinh 2K\sinh K-4\cosh 2K \left(2P_{\boldsymbol{\theta}}-3\cosh K \right)}{\cosh^3 K +1-\sinh^2 K\cdot P_{\boldsymbol{\theta}}} \\
    &-\frac{\sinh^2 2K \left(2P_{\boldsymbol{\theta}}-3\cosh 2K \right)^2}{\left(\cosh^3 K+1-\sinh^2 K \cdot P_{\boldsymbol{\theta}}\right)^2}\Bigg\}.
\end{split}
\label{eq:c_ising}
\end{equation}
The total magnetic specific heat is the sum of the magnon and Ising contributions, $c_\text{M}=c_{\text{mag}}+c_{\text{Is}}$. Following this scheme, the thermodynamic properties of FePS$_3$ are computed and shown in Fig. \ref{fig:fig3}. 
\end{document}